# Hierarchical Route Optimization by Using Tree information option in a Mobile Networks

K.K.Gautam
Department of Computer Science & Technology
Roorkee Engineering & Management Technology Institute
Shamli-247 774 (INDIA)

Menu Chaudhary
Department of Computer Science & Technology
Roorkee Engineering & Management Technology Institute
Shamli-247 774 (INDIA)

*Abstract*—The Networks Mobility (NEMO) Protocol is a way of managing the mobility of an entire network, and mobile internet protocol is the basic solution for Networks Mobility. A hierarchical route optimization system for mobile network is proposed to solve management of hierarchical route optimization problems. In present paper, we study Hierarchical Route Optimization Scheme using Tree Information Option (HROSTIO). The concept of optimization-finding the extreme of a function that maps candidate 'solution' to scalar values of 'quality' – is an extremely general and useful idea. For solving this problem, we use a few salient adaptations and we also extend HROSTIO perform routing between the mobile networks.

I. INTRODUCTION

In the trend of ubiquitous computing, many electric appliances and electronic devices capable of integrating with wireless communications are being added. The mobile internet protocol (IP) working group within the Internet Engineering Task Force (IETF) has proposed the mobile IP protocol [1], [2] to support host mobility in IP based networks. The mobile IP aims at maintaining internet connectivity while a host is moving. The Networks Mobility (NEMO) protocol is a way of managing the mobility of an entire network, viewed as a single unit, which changes its points to attachments in the internet [3]. Such an internet will include one or more Mobile Routers (MRs) that connect it to the global internet. A mobile network can connect it to the global internet.

A mobile network can have a hierarchical structure; in this paper we propose a Hierarchical Route Optimization Scheme using Tree Information Option (HROSTIO) for mobile network. In addition to routing inefficiency, other criteria are important in designing a route optimization scheme for mobile networks. The concepts of network mobility have been introduced to reduce the signaling overheads of a number of hosts moving as group. The NEMO basic support protocol uses a bidirectional tunnel between the Home Agent (HA) and the Mobile Networks Needs (MNNS) from sending all there location registration simultaneously when the MR changes its point of attachment. The characteristic is called mobility transparency, which is a very desirable feature for the route optimization scheme.

Mobile networks can here very complex form of hierarchy e.g. Mobile networks in a mobile network Visiting Mobile Nodes (VMNS) in mobile networks and so on. This situation is repaired as nested mobile network.

Many important problems arising in science, industry and commerce, mobile networks fall very neatly into the read-made category of optimization problem. That is to say, these problems are solved if we can simply find a solution that maximizes or minimizes some important and measurable property.

For example, in Fig. 1, we might want to find the set of mobile router in a simple illustration of nested mobile network at the beginning $MR_1$, $MR_2$, and Visiting. Mobile Nodes (VMNS) are attached to their own home link. After $MR_1$ moves to a foreign link, $MR_2$ moves this make a simplest form of nested mobile networks.

II. NEMO ARCHITECTURE

When a mobile network moves from one place to another, it changes its points of attachment to the internet, which also makes changes to its reach ability and to the Internet topology. NEMO (Network Mobility) working group has come up with NEMO support solution. NEMO support is a mechanism that maintains the continuity of session between Mobile Networks Node (MNN) and their Correspondent Nodes (CN) upon a mobile Router's change of point attachment. NEMO support is divided into two parts:

*1). NEMO Basic Support*
*2). NEMO Extended Support*

NEMO Basic Support is a solution for persevering session continuity by means of bidirectional tunneling between Home Agent (HA) and a mobile network. And NEMO extended Support is a solution for providing the necessary optimization between arbitrary Mobile Networks Nodes and correspondent Nodes, including routing optimization [5]. There has not been much research done with the NEMO extended Support Protocol.

A mobile Network is composed of one or more IP subnets viewed as a single unit. The Mobile Router is the gateway for the communication between the mobile network and the internet.

An Access Router (AN) is a router at the edge of an access network which provides wireless link to mobile nodes. A link is simply a physical medium via which data is transformed between multiple nodes. A Home Link is the link attached to the interface at the Home Agent on which the Home Prefix is configured. Any Link other than Home link is foreign link. NEMO link is the link within the mobile network. A Mobile Router has two interfaces:-
Ingress Interface: The interface of the MR attached to a link inside the mobile network.

Egress interface: The interface of the MR attached to the home link if the MR is at home and to foreign link if it is a foreign network.

NEMO Basic Support protocol is an extension to the Mobile Ip version 6 (MIPv6) [2]. MIPv6 is a version of Internet Protocol (IP) that supports mobile nodes.

### III. MOBILE ROUTERS

A Mobile Router is a router that can change its point of attachment to the network by moving from one link to another. All the Internet traffic to and from the mobile network passes through the Mobile Router. Therefore, Mobile Router has to perform certain operations to be able to support network mobility.

### IV. HROSTIO

For the hierarchical Route Optimization scheme using tree information option (HROSTIO) we use an assistant data structure and call it MNN-CN(mobile network node-corresponding node) list .It is stored at MRs and records the relationship of the MNN-CN.

### V. TREE INFORMATION OPTION

The tree information option Tio [4] avoids routing loops in a nested NEMO Fig. 2 shows the TIO formet in an RA message to prevent a loops MRs performance topology based on various metrics fig.1 is defined information option.

| A=0 | B=8016 | | | C=150031 |
|---|---|---|---|---|
| Type | Length | G | H | Reserved |
| Preference | BTRHRO | | | |
| Tree depth | Tree pref. | | Tree delay | |
| Path digest | | | | |
| Tree information system | | | | |

Figure 1. *BTRHRO*(Boots Time Random for Hierarchical route Optimization system)

### VI. PERSONAL AREA NETWORK

A mobile network can have a hierarchical structure e.g. a mobile network within another mobile network. This situation is referred to as mobile network. A Personal Area Network (PAN) may travel a vehicle, which also contains a mobile network of larger scale MR-1, MR-2, MR-3 … are attaché their own home link. A wireless personal area network moves as a single unit with one or more mobile routers that connect it to global internet.

### VII. EXTENDED TIO

The tree information option for Hierarchical Route Optimization scheme in network mobility is divided in two parts:

*A.* Basic Hierarchical Route Optimization Scheme using Tree Information Option (HROSTIO). This is define hierarchical Route Optimization scheme between the Corresponding Nodes (CNs) and the Local Fixed Nodes (LFNs)

*B.* Extended Hierarchical Route optimization Scheme using Tree Information Option (HROSTIO).

### VII. HIERARCHICAL ROUTE OPTIMIZATION BY PACKES DISRUPTION TREE (HROPD)

For the Hierarchical Route Optimization Scheme using the Edmond's Theorem. The tree system we study a one routing assignment, for information packets to go from the root of the directed tree, if sender is x and receiver is y. then we present a subgraph

$G_k = (V_k, E_k, C_k) \leq G = (V, E, C)$ when

$E_k$ = Edge Set

$V_k$ = Vertex

$C_k$ = Edge Capacity

Hence a distribution tree $G_k = (V_k, E_k, C_k)$ can deliver communication to receiver at a rate

$R(G_k) = \min_{e \in E_k} c_k(e).$

Hence we present hierarchical edge packs, as described below:

- Hierarchical edge bundling is a flexible and generic method that can be used in conjunction with existing tree visualization techniques to enable users to choose the tree visualization that they prefer and to facilitate integration into existing tools.

- Hierarchical edge bundling reduces visual clutter when dealing with large number of adjacency.

- Hierarchical edge bundling provide an intuitive and continuous way to control the strength of bundling. Low Bundling strength mainly provides low-level, node-to-node connectivity information, whereas high bundling strength provides high-level information as well by implicit visual of adjudges between parent nodes that they are the result of explicit adjacency edges between their respective child nodes Hierarchical edge bundling provide an intuitive and continuous way to control the strength of bundling. Low

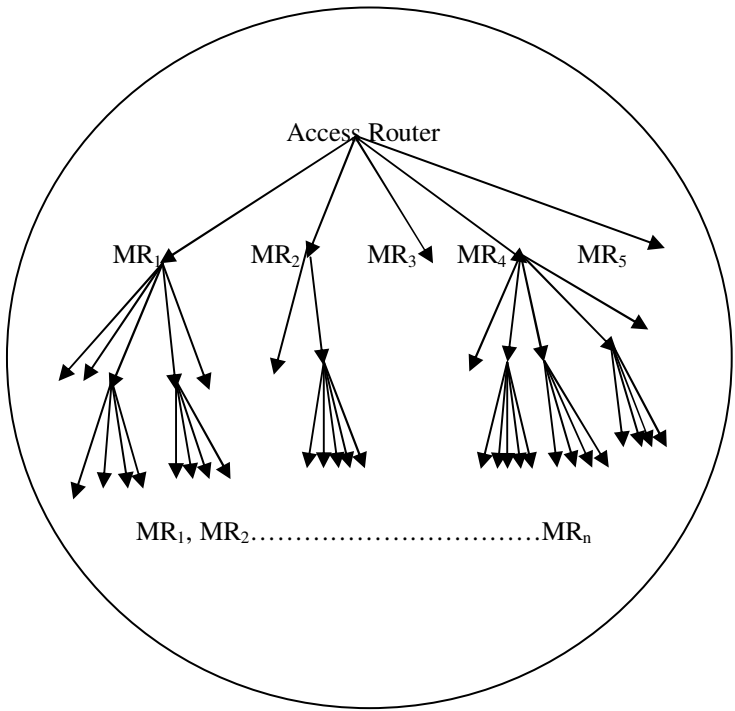

Fig.-2

CONCLUSION

Hierarchical Route Optimization scheme optimization scheme in mobile network modifying the process of Tree information option. And hence the NEMO basic support protocol needs to be extended with an appropriative route optimization scheme. the optimization scheme to easily solved by Tree information option. We propose a scheme can achieve the Hierarchical Route Optimization Scheme using Tree Information Option (HROSTIO) for route optimization environment.

AUTHORS PROFILE

Authors Profile .. K K Gautam is the Dean in the Roorkee Engineering & Management Technology Institute, Shamli-247774, India. Prof Gautam is basically a mathematician and is working in the area of mobile network and wireless network for the past 3 years.

Meenu Chaudhury is a lecturer in the Department of Computer Science at Roorkee Engineering & Management Technology Institute, Shamli-247 774, India. Meenu has B Tech degree in Computer Science to her credit.